\documentclass[submission,copyright,creativecommons]{eptcs}
\providecommand{\event}{MARS 2026} 

\usepackage{iftex}

\ifpdf
  \usepackage{underscore}         
  \usepackage[T1]{fontenc}        
\else
  \usepackage{breakurl}           
\fi

\title{Modelling and Analysis of Supply Chains using\\ Product Time Petri Nets}
\author{Eric Lubat 
\institute{IRIT}
\institute{Université Toulouse\\
Toulouse, France}
\email{eric.lubat@irit.fr}
\and 
Pierre-Emmanuel Hladik
\institute{Nantes Université, Ecole Centrale Nantes, CNRS, LS2N,}
\institute{UMR 6004, 44000, Nantes, France}
\email{pierre-emmanuel.hladik@ls2n.fr}
\and 
Yoann Mateu\\
Rémi Sauvère
}

\def\authorrunning{Lubat, Hladik, Mateu, Sauvère}
\def\titlerunning{Modelling and Analysis of Supply Chains using Product Time Petri Nets}

\usepackage[utf8]{inputenc}
\usepackage{hyperref}
\usepackage{xcolor}
\usepackage{mathrsfs}
\usepackage[T1]{fontenc}
\usepackage{pdfpages}
\usepackage{algorithm}
\usepackage{graphicx}
\usepackage{latexsym}
\usepackage{textcomp}
\usepackage{amsmath,amsfonts,amstext,amssymb,verbatim}
\usepackage{stmaryrd}
\usepackage[nointegrals]{wasysym}
\usepackage{wrapfig}
\usepackage{listings}
\usepackage{color}
\usepackage{footnote}
\usepackage{multirow} 
\usepackage{array}
\usepackage{xspace}
\usepackage{subfig}
\usepackage{afterpage}
\usepackage{float}

\usepackage{alltt}
\usepackage{listings}
\usepackage{placeins}
\usepackage{calculator}
\usepackage{siunitx}
\usepackage{adjustbox}

\DeclareOldFontCommand{\bf}{\normalfont\bfseries}{\mathbf}

\hypersetup{
  colorlinks   = true, 
  urlcolor     = blue, 
  linkcolor    = blue, 
  citecolor   = red 
}
\definecolor{blue}{RGB}{0, 102, 204}
\definecolor{brick-red}{RGB}{203, 65, 84}
\definecolor{brown}{RGB}{216, 203, 175}
\definecolor{green}{RGB}{0, 153, 76}
\definecolor{grey}{RGB}{151, 100, 100}
\definecolor{orange}{RGB}{255, 166, 48}
\definecolor{purple}{RGB}{102, 0, 204}
\definecolor{red}{RGB}{202, 3, 3}

\usepackage{siunitx}
\usepackage{booktabs} 

\usepackage{tabularx}

\newcolumntype{Y}{>{\hfill\arraybackslash}X}

\newcolumntype{C}{>{${}}c<{{}$}}
\newcolumntype{L}{>{${}}l<{{}$}}
\newcolumntype{R}{>{${}}r<{{}$}}

\newcommand{\vect}[1]{\ensuremath{\mathbf{#1}}}

\newcommand{\Lab}{\ensuremath{\mathcal{L}}\xspace}

\newcommand{\TPN}[0]{{TPN}\xspace}
\newcommand{\TTPN}[0]{{PTPN}\xspace}

\newcommand*{\dotgeq}{\mathrel{\dot{\geqslant}}}

\newcommand{\SG}{{{S}}}

\newcommand{\pRat}{\ensuremath{\mathbb{Q}_{\ge 0}}\xspace}

\newcommand{\Itrv}{\ensuremath{\mathbb{I}\xspace}}

\makeatletter
\newcommand{\@dotminus}{%
  \ooalign{\hidewidth\raise1ex\hbox{.}\hidewidth\cr$\m@th-$\cr}%
}
\makeatother

\makeatletter
\def \rightarrowfill{\m@th\mathord{\smash-}\mkern-6mu%
  \cleaders\hbox{$\mkern-2mu\mathord{\smash-}\mkern-2mu$}\hfill
  \mkern-6mu\mathord\rightarrow}
\makeatother

\makeatletter
\def \Rightarrowfill{\m@th\mathord{\smash-}\mkern-6mu%
  \cleaders\hbox{$\mkern-2mu\mathord{\smash-}\mkern-2mu$}\hfill
  \mkern-6mu\mathord\Rightarrow}
\makeatother

\makeatletter
\def \rightarrowfill{\m@th\mathord{\smash-}\mkern-6mu%
  \cleaders\hbox{$\mkern-2mu\mathord{\smash-}\mkern-2mu$}\hfill
  \mkern-6mu\mathord\rightarrow}
\makeatother

\makeatletter
\def \Rightarrowfill{\m@th\mathord{\smash=}\mkern-6mu%
  \cleaders\hbox{$\mkern-2mu\mathord{\smash=}\mkern-2mu$}\hfill
  \mkern-6mu\mathord\Rightarrow}
\makeatother

\makeatletter
\def \midrightarrowfill{\m@th\mathord{\smash{\raisebox{.2ex}{$\scriptscriptstyle\mid$}}\!\!\,-}\mkern-6mu%
  \cleaders\hbox{$\mkern-2mu\mathord{\smash-}\mkern-2mu$}\hfill
  \mkern-6mu\mathord\rightarrow}
\makeatother

\makeatletter
\def \midRightarrowfill{\m@th\mathord{\smash{\raisebox{.1ex}{$\scriptstyle\mid$}}\!\!\!=}\mkern-6mu%
  \cleaders\hbox{$\mkern-2mu\mathord{\smash=}\mkern-2mu$}\hfill
  \mkern-6mu\mathord\Rightarrow}
\makeatother

\newcommand{\overstackrel}[2]{\mathrel{\mathop{#1}\limits^{#2}}}
\newcommand{\trans}[1]{\mathbin{\smash[t]{\overstackrel{\rightarrowfill}{\ #1\ }}}}
 
\newcommand{\interp}[1]{{[\![} {#1} {]\!]}}

\newcommand{\tI}[1][]{\ensuremath{\varphi_{#1}}}

\newcommand{\sI}[1]{\ensuremath{x_{#1}}}
\newcommand{\ssI}[1]{\ensuremath{y_{#1}}}

\newcommand{\SIF}{\ensuremath{{\bf I}_s}\xspace}

\usepackage[T1]{fontenc}

\usepackage{tikz}
\usetikzlibrary{arrows}
\usetikzlibrary{decorations.markings}
\usetikzlibrary{shadows}

\tikzset{
  big stealth/.style={
    decoration={markings,mark=at position -(0.1pt) with {\arrow[scale=2*\scale]{stealth}}},
    postaction={decorate},
    shorten >=0.4pt}}
\tikzset{
  big ring/.style={
    decoration={markings,mark=at position -(0.1pt) with {\arrow[scale=1.5*\scale]{o}}},
    postaction={decorate},
    shorten >=8pt*\scale}}
\tikzset{
  big disc/.style={
    decoration={markings,mark=at position -(0.1pt) with {\arrow[scale=1.5*\scale]{*}}},
    postaction={decorate},
    shorten >=8pt*\scale}}
\tikzset{
  big box/.style={
    decoration={markings,mark=at position -(0.1pt) with {\arrow[scale=1.5*\scale]{open square}}},
    postaction={decorate},
    shorten >=8pt*\scale}}
\tikzset{
  big tile/.style={
    decoration={markings,mark=at position -(0.1pt) with {\arrow[scale=1.5*\scale]{square}}},
    postaction={decorate},
    shorten >=8pt*\scale}}

\tikzstyle{state}=[circle, very thick, fill, top color=white, bottom color=white, draw=black, minimum size=40pt, drop shadow]
\tikzstyle{place}=[circle, very thick, fill, top color=white, bottom color=white, draw=black, minimum size=40pt, drop shadow]
\tikzstyle{trans}=[rectangle, very thick, fill, top color=white, bottom color=white, draw=black, minimum size=32pt, drop shadow]
\tikzstyle{arc}=[thick, big stealth, black]
\tikzstyle{read}=[thick, big disc, black]
\tikzstyle{inhibitor}=[thick, big ring, black]
\tikzstyle{stopwatch}=[thick, big tile, black]
\tikzstyle{stopwatchinhibitor}=[thick, big box, black]
\tikzstyle{priority}=[thick, big stealth, orange]
\tikzstyle{enabling}=[thick, big disc, orange]
\tikzstyle{disabling}=[thick, big ring, orange]
\tikzstyle{token}=[circle, fill, draw=black, minimum size=4pt]
\tikzstyle{glob-options}=[label distance=6pt*\scalenodes*\scale,x=1pt,y=-1pt,scale=\scale,every node/.style={transform shape}]
\tikzstyle{virtual}=[circle, draw=white, minimum size=20pt]

\usepackage{rotating}

\usepackage{amsmath,amsfonts,bm,bbm,ifthen,mathtools}
\usepackage{amssymb}

\usepackage{tikz}
\usetikzlibrary{arrows, automata, shapes, calc, petri, arrows.meta}

2

\def\env@cases#1{%
  \let\@ifnextchar\new@ifnextchar
  \left\lbrace
  \def\arraystretch{1.2}%
  \array{@{}#1@{\quad}l@{}}%
}

\newcommand*\dotminus{\buildrel\textstyle.\over{%
    \hbox{\vrule height3pt depth0pt width0pt}{\smash-}}}

\renewcommand*\epsilon{\varepsilon}
\renewcommand*\phi{\varphi}

\newcounter{ex@mples@ve}

\newcommand*\bbot{%
  \hbox{%
    \hbox to -.475pt{%
      \raisebox{1.5pt}[0pt][0pt]{%
        \rule{6.4pt}{.4pt}%
      }%
      \hss}%
    $\bot$%
  }}
\newcommand*\ttop{%
  \hbox{%
    \hbox to -.475pt{%
      \raisebox{5pt}[0pt][0pt]{%
        \rule{6.4pt}{.4pt}%
      }%
      \hss}%
    $\top$%
  }}

\newcommand*\modaldiamond[2]{%
  \ifthenelse{\equal{#1}{}}{%
    \ensuremath{\hspace{0.2ex}\mathchoice{\raisebox{.15ex}{$\scriptstyle \mathopen|$}}{\raisebox{.15ex}{$\scriptstyle \mathopen|$}}{\raisebox{.1ex}{$\scriptscriptstyle \mathopen|$}}{\raisebox{.1ex}{$\scriptscriptstyle \mathopen|$}}\mathclose\rangle\hspace{0.2ex}#2}}{%
    \ensuremath{\mathopen|#1\mathclose\rangle#2}}}

\makeatother
\newcommand\pomsetwop[4]{
  \vcenter{\xymatrix@1@R=#1@C=#2@M=#3{#4}}%
}

\makeatletter

\newbox\dotrightarrow@box
\setbox\dotrightarrow@box\hbox
{%
  \begin{tikzpicture}
    \draw[dotted,->,>=stealth] (0,0) -- (1.5em,0);
  \end{tikzpicture}%
}
\newcommand*\dotrightarrow
{\relax\ifmmode\expandafter\dotrightarrow@m\else\expandafter\dotrightarrow@t\fi}
\newcommand*\dotrightarrow@t[1][1.5em]
{\resizebox{#1}{!}{\raisebox{.5ex}{\usebox\dotrightarrow@box}}}
\newcommand*\dotrightarrow@m[1][]
{%
  \if\relax\detokenize{#1}\relax
  \mathchoice
  {\dotrightarrow@t}
  {\dotrightarrow@t}
  {\dotrightarrow@t[1.1em]}
  {\dotrightarrow@t[0.9em]}%
  \else
  \dotrightarrow@t[#1]%
  \fi
}

\renewcommand\thefootnote{\@arabic\c@footnote}

\newtheorem{mydef}{Definition}

\begin{document}
\maketitle

\begin{abstract}
Supply chains involve geographically distributed manufacturing and assembly sites that must be coordinated under strict timing and resource constraints. While many existing approaches rely on Colored Petri Nets to model material flows, this work focuses on the temporal feasibility of supply chain processes. 
We propose a modular modelling approach based on Product Time Petri Nets (PTPNs), where each subsystem is represented independently and the global behaviour emerges through synchronised transition labels. A key feature of the model is the explicit representation of the supply chain manager as a critical shared and mobile resource, whose availability directly impacts system feasibility.
We analyse how timing constraints and managerial capacity influence the system behaviour, identifying configurations that lead to successful executions, timeouts, or timelocks induced by incompatible timing constraints. This approach enables systematic \emph{what-if} analysis of supply chain coordination policies and demonstrates the relevance of PTPNs for modelling and analysing synchronised timed systems.
\end{abstract}

\section{Introduction}

Supply chains involve a geographically distributed network of manufacturing and assembly sites. Key components such as electronic components are supplied by subcontractors, while final assembly is carried out in a factory. Coordination between these sites is tightly constrained by just-in-time logistics, strict quality control procedures, and shared critical resources. The role of the supply chain manager is essential in resolving these tightly constrained processes, particularly when quality deviations occur.

Understanding how local timing deviations, such as a delay in a supplier site or a late modification propagate through the supply chain, is crucial to prevent bottlenecks, minimizing downtime, and optimizing throughput. However, modelling such complex interactions often leads to monolithic, non-modular representations that are difficult to analyse, simulate, or reuse.

In this work, our contribution is a benchmark based on a modular approach using \emph{Product Time Petri Nets} (PTPNs)~\cite{LubatDBPS19}. Each subsystem (e.g., a manufacturing site, a transport leg, or a quality control process) is represented by a separate \emph{Time Petri Net} (TPN)~\cite{merlin1974study}, and synchronisation between subsystems is enforced through shared transition labels. This modular structure allows engineers to explore the global impact of local timing variations while preserving model modularity and supporting compositional analysis.

Notably, the supply chain manager is explicitly modelled as a critical mobile resource, moving from supplier to supplier to process non-conformity acceptance and approve the delivery of critical parts. This modelling choice contrasts with more classical approaches based on TPNs or CTPNs (Colored Time Petri Nets)~\cite{LIU2007761}, in which the supply chain is typically represented as a monolithic model.

The proposed PTPN-based benchmark is intended to support analysis of supply-chain coordination policies. By varying parameters such as the number of suppliers, the number of managers, or timing bounds on production and validation activities, one can evaluate feasibility, detect deadlocks, and study the conditions under which coordination becomes impossible. Rather than aiming at the faithful reproduction of a specific industrial system, the model provides a structured and parametric case study suitable for evaluating modelling and analysis techniques for synchronised timed systems. Our approach has been implemented in a
tool called TWINA~\cite{LubatDBPS19} (from one of the authors' thesis) and verified using the TINA toolbox~\cite{berthomieu2004}, with the tool \emph{selt}~\cite{BPV07}, a state-event LTL (SE-LTL) checker. Using this model checker, we examine the states and transitions that lead to infeasible supply-chain configurations.

The remainder of the paper is organized as follows. Section~\ref{sec:rew} reviews related work on supply-chain modelling with Petri Nets and timed formalisms. Section~\ref{sec:tpn} recalls the definition and semantics of Time Petri Nets and Product Time Petri Nets. Section~\ref{sec:sc} presents the proposed supply-chain benchmark and its modular structure (available online\footnote{\url{https://github.com/Darkelubat/SupplyChain}}\footnote{\url{https://zenodo.org/records/18925093}}). Section~\ref{sec:sAa} reports experimental results obtained through model checking. Finally, Section~\ref{sec:conclusion} concludes and discusses perspectives for future work.

\section{Related Work}
\label{sec:rew}

The modelling and analysis of supply chains have long been a topic of interest in both the
modelisation and formal methods communities. In particular, Petri Nets and their timed
extensions have been widely used to represent concurrency, synchronisation, and resource
constraints in manufacturing and logistics systems~\cite{Zhang2009}. Their formal semantics and graphical nature
make them suitable to capture the complex interactions that arise in distributed industrial
processes.

Time Petri Nets (TPN)~\cite{merlin1974study,BPV06} extend classical Petri Nets by associating temporal constraints with transitions, enabling the representation of processing times, transport delays, and waiting periods. As a result, they have been successfully applied to workflow modelling, performance evaluation, and the verification of time-critical systems.

Several works by van der Aalst and collaborators~\cite{vanderAalst1994ModellingAA,WorkflowAalst} address the modelling of supply chains and logistics processes using Petri Nets and workflow formalisms. These approaches focus on the correlation of events, the timing constraints between activities, and compliance with predefined temporal patterns. While such models effectively capture global behavioural constraints, they are typically expressed as monolithic nets, which can limit modularity and reuse when modelling large, geographically distributed supply chains. In particular, Bevilacqua et al.~\cite{Mazzuto15082012} employ Time-Coloured Petri Nets to model supply chain processes, where transition firing intervals encode production times, transportation delays, and managerial decision durations, allowing the analysis of system performance and coordination effects.

Modularity and compositionality have been recurrent challenges in Petri Nets modelling, especially in a timing context~\cite{berthomieu1983enumerative}. Various
approaches have been proposed to compose Petri Nets through shared places, transition fusion, or
synchronisation labels. In the context of timed systems, however, synchronisation across multiple
timed components often leads to significant state-space growth~\cite{refIPTPN,cassez2006structural}, which complicates analysis.

In contrast to existing approaches, this work builds on Product Time Petri Nets~\cite{LubatDBPS19,lubat:tel-03528121}, which
enables the synchronised composition of multiple TPNs through shared transition labels. This allows
each supplier, transport process, or decision authority to be modeled independently, while still
capturing their temporal interactions. The proposed framework is particularly suited to study
delay propagation and managerial resource contention in supply chains, as it preserves modularity at the
modelling level and supports systematic experimentation with timing parameters. 

Our approach is closely related to classical scheduling techniques such as the Program Evaluation and Review Technique (PERT)~\cite{malcolm1959pert}, which models a project as a network of activities annotated with execution durations. In PERT analysis, the overall project duration is typically derived from the critical path, computed using conservative estimates of activity durations. Previous work on PTPN has already modelled factory systems in the context of diagnosability analysis for Time Petri Nets~\cite{lubatCDC2020}. However, to the best of our knowledge, PTPN have not yet been applied to the analysis of supply-chain coordination and feasibility.

In this work, we exploit this compositional modelling framework to analyse
the temporal feasibility of supply-chain coordination policies and to study
the impact of timing constraints and managerial resource availability on the emergence
of deadlocks.

\section{Time Petri Nets and Product Time Petri Nets}
\label{sec:tpn}

We describe our model, the TPN, and its extension, the PTPN, in the following section. For a more detailed description of the proof, semantics, execution and languages of PTPN, please refer to its introduction paper~\cite{LubatDBPS19}. 
These definitions and semantic notions justify the use of SCG-based (State Class Graph~\cite{berthomieu1983enumerative,BD91}) verification with TINA and \emph{selt} in Section~\ref{sec:sAa}. This section constitutes the technical core that underpins the contribution of this paper, as described in Section~\ref{sec:sc}.

\subsection{Time Petri Net (TPN) and Product Time Petri Net (PTPN)}

A {\em Time Petri Net} (TPN)~\cite{merlin1974study} is a net where each transition, $t$, is
decorated with a (static) time interval $\SIF(t)$ that constrains the
time at which it can fire. A transition is enabled when there are
enough tokens in its input places. Once enabled, transition $t$ can
fire if it stays enabled for a duration $\theta$ that is in the
interval $\SIF(t)$. In this case, $t$ is said \emph{time enabled} (we refer the reader to~\cite{LubatDBPS19,BPV06} for details). More formally:

\begin{mydef}
A TPN is a tuple $\langle {P},{T},{\vect{Pre}},{\vect{Post}},m_0,\SIF \rangle$ in
which: $\langle {P},{T},{\vect{Pre}},{\vect{Post}} \rangle$ is a net (with ${P}$
and ${T}$ the set of places and transitions);
${\vect{Pre}},~ {\vect{Post}} : {T} \rightarrow {P} \rightarrow \mathbb{N}$ are the
precondition and postcondition functions; $m_0 : P \rightarrow \mathbb{N}$
is the initial marking; and $\SIF : {T} \rightarrow \Itrv$ is the
\emph{static interval function}. We use $\Itrv$ for the set of all
possible time intervals. 
\end{mydef}

We consider that transitions can be tagged using a countable set of
labels, $\Sigma = \{a, b, \dots\}$. We also distinguish the special
constant $\epsilon$ (not in $\Sigma$) for internal, silent
transitions. In the following, we use a global labeling function
$\Lab$ that associates a unique label in $\Sigma \cup \{\epsilon\}$ to
every transition. 

The alphabet of a net is the collection of labels (in $\Sigma$)
associated to its transitions.

In general terms, the semantics of a
TPN is a TTS structure $\langle S,S_0,\rightarrow\rangle$ with only two possible kinds of actions: either
a transition $t$ is fired, or a time delay $\theta$ elapses. A transition $t$ can fire from the
state $(m, \tI)$ if $t$ is enabled at $m$ and firable instantly. 

We use an extension of TPN in which it is possible to fire
several transitions ``{synchronously}''. A {Product TPN}~\cite{LubatDBPS19} (PTPN) is the
composition $(N, R)$ of a net $N$, with transitions $T$, and a
\emph{(product) relation}, $R$, that is a collection of \emph{firing
  sets} $r_1, \dots, r_n$ included in $T$ (hence
$R \subseteq \mathcal{P}(T)$, the powerset of $T$). The idea is that
all the transitions in an element $r$ of $R$ should be fired at the
exact same time. As a consequence, two transitions in $r$ should have
the same labels (we should use $\Lab(r) = a$ to say they have a common
label $a$) and not interfere with each other (they should not share a
common input place).

\begin{mydef}
A Product TPN $(N, R)$ is the pair of a net $N = \langle {P},{T},{\vect{Pre}},{\vect{Post}},m_0,\SIF \rangle$ and a product
relation $R \subseteq \mathcal{P}(T)$ such that, for every firing
set $r$ in $R$, transitions in $r$ are independent and compatible:
if both $t_1$ and $t_2$ are in $r$ then $\Lab(t_1) = \Lab(t_2)$ and
for every place $p$ in $P$,
$\vect{Pre}(t_1)(p) > 0 \Rightarrow \vect{Pre}(t_2)(p) = 0$.
\end{mydef}

A firing set is a set of transitions that must fire together (e.g., $\{t_1, t_2\}$ with $\Lab(t_1) = \Lab(t_2)$). In a TPN, transitions are not merged; rather, they must fire synchronously in accordance with their timing constraints.

Next, we summarize the behaviour of nets. Apart from the effect of the
firing sets, the following definitions are quite standard, see for
instance~\cite{berard2005comparison,BPV06}.

A \emph{marking} $m$ of a net $\langle {P},{T},{\vect{Pre}},{\vect{Post}} \rangle$
is a mapping $m : {P} \rightarrow \mathbb{N}$ from places to natural
numbers. A transition $t$ in ${T}$ is \emph{enabled} at $m$ if and
only if $m \dotgeq \vect{Pre}(t)$, where $\dotgeq$ is the pointwise
comparison between functions. Pointwise comparison in Petri net markings refers to comparing two markings, place by place over the set of places.

A \emph{state} of a {\TTPN} is a pair $s = (m, \tI)$ in which $m$ is a
marking, and $\tI: {T} \to \Itrv$ is a mapping from transitions to
time intervals, also called \emph{firing domains}. Intuitively, if $t$
is enabled at $m$, then $\tI(t)$ contains the dates at which $t$ can
possibly fire in the future. For instance, when $t$ is newly enabled,
it is associated to its static time interval $\tI(t) =
\SIF(t)$. Likewise, a transition $t$ can fire immediately only when
$0$ is in $\tI(t)$ and it cannot remain enabled for more than its
timespan, {\it i.e.} the maximal value in $\tI(t)$. We represent the state in a SCG structure
like in~\cite{berthomieu1983enumerative}.

The semantics of a \TTPN is a (labelled) Kripke structure, or Time
Transition System (TTS), $\langle S,S_0,\rightarrow\rangle$, with two
possible kinds of actions: either $s \trans{a} s'$, meaning that a set
of transitions $t$ with label $a$ is fired from $s$; or
$s \trans{\theta} s'$, with $\theta \in \pRat$ (where $\pRat$ denotes the set of non-negative rational numbers), meaning that we let a
duration $\theta$ elapse from $s$. A transition $t$ can fire from
state $(m,\tI)$ if $t$ is enabled at $m$ and firable instantly. When
we fire a set of transitions $r = \{t_1, \dots, t_n\}$ from state
$(m, \tI)$, a transition $k$ (with $k \neq t$) is said to be
\emph{persistent} if $k$ is also enabled in the marking
$m - \sum_{t \in r}\vect{Pre}(t)$, that is if
$m - \sum_{t \in r}\vect{Pre}(t) \dotgeq \vect{Pre}(k)$. The other transitions
enabled after firing $r$ are called \emph{newly enabled}.

\begin{mydef}
  The semantics of a {\TTPN} can be formally defined as $(N, R)$, with
  $N = \langle {P}, {T}, {\vect{Pre}}, {\vect{Post}}, m_0, \SIF \rangle$, is the
  TTS ${\langle \SG, s_0, \trans{} \rangle}$, also denoted
  $\interp{(N, R)}$, where $S$ is the smallest set containing $s_0$
  and closed by $\trans{}$ such that:\par

  \noindent --- the initial state is $s_0 = (m_0, \tI[0])$ where
  $\tI[0]$ is the firing domain such that $\tI[0](t) = \SIF(t)$ for
  every $t$ enabled at $m_0$;

  \noindent --- the state relation
  ${\rightarrow} \subseteq S \times ({{\Sigma} \cup \{\epsilon\}
    \cup \pRat})\! \times S$ is such that for all state $(m, \tI)$
  in $\SG$
  \begin{itemize}
  \item[(i)] if $r \in R$ with labels $a$ and $t$ is enabled at $m$
    and $0 \in \tI(t)$ for all $t \in r$, then
    $(m, \tI) \trans{a} (m', \tI')$ where
    $m' = m - \sum_{t \in r} \vect{Pre}(t) + \sum_{t \in r} \vect{Post}(t)$ and
    $\tI'$ is a firing function such that $\tI'(k) = \tI(k)$ for any
    persistent transition and $\tI'(k) = \SIF(k)$ elsewhere.
    
  \item[(ii)] if $\tI(t) - \theta$ is defined for all $t$ enabled at
    $m$ then $(m,\tI) \trans{\theta} (m,\tI \dotminus \theta)$.
     $\tI(t) - \theta$ denotes the interval obtained by subtracting $\theta$ from both bounds.
  \end{itemize}
\end{mydef}

Transitions in the case $(i)$ above are called \emph{discrete}; those
labelled with delays (case $(ii)$) are the \emph{continuous}, or time
elapsing, transitions.

A Product TPN $(N, R)$ allows to fire multiple transitions
simultaneously, constrained by the relation $R$. Therefore \TPN form a
natural subset of \TTPN, the one where every firing set has only one
transition. More precisely, we can always interpret a \TPN $N$ with
transitions $\{t_1, \dots, t_n\}$ as the \TTPN $(N, R_N)$, where $R_N$
is the collection of singleton $\{\{t_1\}, \dots, \{t_n\}\}$. In the
following, we often omit the product relation in a \TTPN when it is
not needed, or obvious from the context. We should also simply use the
term net, or the symbol $N$, to refer to a Product TPN.

\subsection{Synchronous Product of PTPN} 

We can define the product of two PTPN as follows:

\begin{mydef}
Given two nets $(N_1, R_1)$ and $(N_2, R_2)$ with disjoint sets of
places $P_1, P_2$ and transitions $T_1, T_2$, their product
$(N_1, R_1) \times_L (N_2, R_2)$ is the \TTPN $(N, R)$ where $N$ is
the concurrent composition (juxtaposition) of $N_1$ with $N_2$, the
net
$\langle {P_1} \cup P_2, {T_1 \cup T_2}, {\vect{Pre}}, {\vect{Post}}, m^1_0 \uplus
m^2_0, \SIF \rangle$ with $\vect{Pre}(t)(p) = \vect{Pre}_i(t)(p)$ if and only if
$t \in T_i$ and $p \in P_i$ with $i \in 1..2$, and $0$ otherwise (same
with $\vect{Post}$); and the product relation $R$ is such that:
\[ R =
  \begin{array}[t]{l}
    \displaystyle  \bigcup_{a \in L} \{ r_1 \cup r_2 \mid r_i \in R_i, \Lab(r_i) = a,
    i \in 1..2\}\\
    \displaystyle \cup  \bigcup_{a \in \Sigma \setminus L \cup \{
    \epsilon \}} \{ r \mid r \in R_1 \cup R_2 ,  \Lab(r) = a  \} 
  \end{array}
\]
\end{mydef}

Unlike the conventional \emph{synchronous composition} operator
between Petri Nets, we do not merge transitions with the same labels
but compose relations instead. But as with synchronisation, our
goal is to define an operation that is a \emph{congruence}, meaning
that $\interp{N_1 \times_L N_2}$ is equivalent to
$\interp{N_1} \mathbin{\|_L} \interp{N_2}$ where $\mathbin{\|_L}$ denotes the classical synchronous product on labels for TPNs.

\subsection{State Class Graph}

In this subsection, we would like to remind the result on the \emph{state class abstraction} method for TPNs, as defined by Berthomieu et al.~\cite{berthomieu1983enumerative}. A \emph{State Class Graph} (SCG) is a finite abstraction of the timed transition system (TTS) of a net that preserves its markings and traces. The construction is based on the idea that temporal information in states (the firing domain $\tI$) can be conveniently represented using systems of difference constraints~\cite{Ramalingam95solvingdifference}.

\begin{mydef}
A state class $C$ is defined by a tuple $(m, D)$, where $m$ is a marking and the firing domain $D$ is described by a (finite) system of linear inequalities.

In a domain $D$, we use variables $\sI{t}, \ssI{t}, \dots$ to denote a
constraint on the value of $\tI(t)$. A domain $D$ is defined by a set
of difference constraints in reduced form, that is inequalities of the
kind: $\alpha_i \leq \sI{i}$, $\sI{i} \leq \beta_i$ and
$\sI{i} - \sI{j} \leq \gamma_{i,j}$, where $i, j$ range over a given
subset of ``enabled transitions'' and the coefficients $\alpha, \beta$
and $\gamma$ are rational numbers.
\end{mydef}

For a more detailed description of SCG in PTPN, please refer to its original publication~\cite{LubatDBPS19}

\subsection{Example of PTPN and timelock}

\begin{figure}[h]
    \centering
          \def\scale{0.75}
      \def\scalenodes{0.75}
      \raisebox{-0.5\height}{\begin{tikzpicture}[glob-options]
        \node[place, label=-90:{\large $p_{1}$}](p0) at (100.0, 200.0) {};
        \node[place, label=-90:{\large $p_{2}$}](p1) at (200.0, 200.0) {};
        \node[place, label=90:{\large $p_{0}$}](p2) at (100.0, 80.0) {};
        \node[token] at (p2) {};
        \node[trans, label=0:{\large \bf a}, label=-180:{$[2,4]$}](t3) at (100.0,130.0) {\large $t_{0}.1$};
        \node[trans, label=-90:{\large \bf b}, label=-180:{$[1,2]$}](t4) at (25.0,130.0) {\large $t_{1}.1$};
        \node[trans, label=0:{\large \bf b}, label=90:{$[3,4]$}](t5) at (200.0,80.0) {\large $t_{3}.1$};
        \node[trans](t6) at (150.0,200.0) {\large $t$};
        \draw[arc](t4) -- (p2);
        \draw[arc](p0) -- (t4);
        \draw[arc](t5) -- (p2);
        \draw[arc](p1) -- (t5);
        \draw[arc](t6) -- (p1);
        \draw[arc](p0) -- (t6);
        \draw[arc](t3) -- (p0);
        \draw[arc](p2) -- (t3);
      \end{tikzpicture}}
    \qquad{\Huge$\times$}\qquad
    \raisebox{-0.5\height}{\begin{tikzpicture}[glob-options]
        \node[place, label=-90:{\large $p_{1}$}](p0) at (100.0, 200.0) {};
        \node[place, label=90:{\large $p_{0}$}](p2) at (100.0, 80.0) {};
        \node[token] at (p2) {};
        \node[trans, label=0:{\large \bf a}, label=-180:{$[2,4]$}](t3) at (100.0,130.0) {\large $t_{0}.2$};
        \node[trans, label=-90:{\large \bf b}, label=-180:{$[1,2]$}](t4) at (25.0,130.0) {\large $t_{1}.2$};
        \draw[arc](t4) -- (p2);
        \draw[arc](p0) -- (t4);
        \draw[arc](t3) -- (p0);
        \draw[arc](p2) -- (t3);
      \end{tikzpicture}}
    \caption{$N_1$ and $N_2$ before their product}
    \label{fig:Example}
\end{figure}
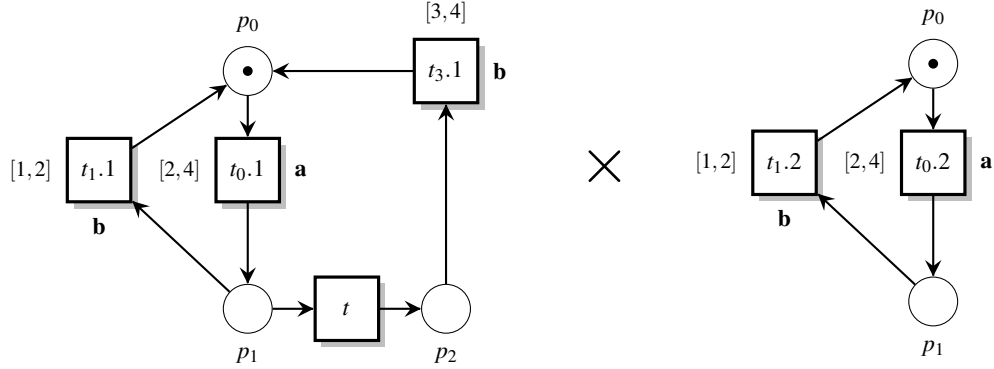

\begin{figure}[h]
    \centering
\def\scale{0.75}
\def\scalenodes{0.75}
\begin{tikzpicture}[glob-options]
\node[state](s0) at (35.0, 30.0) {\large $_{1}$};
\node[state](s1) at (125.0, 30.0) {\large $_{2}$};
\node[state](s2) at (35.0, 115.0) {\large $_{0}$};
\draw[arc](s2) .. controls +(57.2:36.9pt) and +(301.0:42.2pt) .. (s0);
\path(s0)+(301.0:21.1pt) node [label=0:{\large $t_{0}.1|t_{0}.2$}] {};
\draw[arc](s0) -- (s2);
\path(s2)+(90.0:20.0pt) node [label=180:{\large $t_{1}.1|t_{1}.2$}] {};
\draw[arc](s0) -- (s1);
\path(s1)+(180.0:20.0pt) node [label=90:{\large $t$}] {};
\end{tikzpicture}
    \caption{SCG of the PTPN of $N_1$ and $N_2$}
    \label{fig:Example2}
\end{figure}
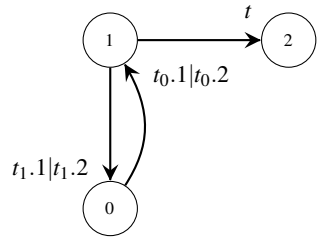

We consider the TPNs $N_1$ and $N_2$ shown in Figure~\ref{fig:Example}, together with the SCG of the resulting PTPN shown in Figure~\ref{fig:Example2}.

Both TPNs share the same alphabet, $\{a,b\}$. Therefore, in the resulting PTPN, transitions labelled with the same symbol are synchronized. The PTPN can thus be simplified into groups of transitions that must fire simultaneously. In this example, we obtain the following groups:
\[
\{\{t_{0.1}, t_{0.2}\}, \{t_{1.1}, t_{1.2}\}, \{t_{3.1}, t_{1.2}\}, \{t\}\}.
\]

All transitions within a given group must fire simultaneously. However, the group $\{t_{3.1}, t_{1.2}\}$ has an empty firing domain: $t_{3.1}$ can fire only after $3$ time units, whereas $t_{1.2}$ must fire before $2$ time units. Hence, these timing constraints are incompatible, which leads to a \emph{timelock}.

More generally, a \emph{timelock} induced by synchronisation occurs when the timing constraints of two or more synchronized transitions admit no common solution. In that case, the system reaches a temporal deadlock.

Firing the unlabelled transition $t$ leads to a dead-end state, illustrating the timelock caused by incompatible timing constraints.

\section{Supply Chain}
\label{sec:sc}

Now that we have introduced the technical background on TPN and PTPN, we introduce the studied system: a distributed network of manufacturing and assembly sites, namely a supply chain. In this paper, the final assembly line, corresponding to the main factory, is called $BAZ$ (for Baziège, a
town in Occitanie). Time units in the model represent days, reflecting the long time scales typically involved in supply chain processes. Our confidence in the model is supported both by prior research in the field and by the operational experience of one of the authors, who has worked as a supply chain engineer since 2019. This ensures that the model captures representative timing constraints and coordination behaviours commonly observed in practice.

A supply chain represents the set of flows required to ensure product delivery. A product may be of various nature, such as a physical object, a service, or an information. Consequently, a supply chain may involve a wide range of activities such as manufacturing, shipping, maintenance operations, and financial transactions. A supply chain encompasses the combination of all resources and means (machines, workforce, information channels, financial flows, transportation, and quality management) that contribute to the execution of industrial processes. Poor coordination or mismanagement of any of these elements may directly impact process lead times. 



Every actor in the supply chain answers to a customer request. For example, suppliers deliver products to manufacturers in response to purchase orders. All actors are interconnected and must be coordinated in order to satisfy the final customer's demands. More specifically, raw material quality and delivery times must be managed so that the overall flow, from raw material production to product delivery to the final customer meets the required time constraints. In addition to material flows, supply chains also involve financial flows (payments, contracts) and data flows (orders, acknowledgements, quality reports), which coordinates and constrains physical production and delivery processes.
This strong coupling between concurrent activities, causal relations, and timing constraints naturally motivates the use of a Time Petri Net.

A supply chain induces a global lead time, defined as the sum of the lead times of all supply-chain actors, from suppliers to delivery to the final customer. Each lead time is characterized by a best-case and a worst-case duration, representing the minimum and maximum time required to perform the corresponding operation. In this work, only worst-case durations are considered, since early completion does not reduce the start time of subsequent operations. This assumption is consistent with a worst-case analysis of distributed industrial processes.

\begin{figure}[h]
    \centering
    \includegraphics[width=1.0\linewidth]{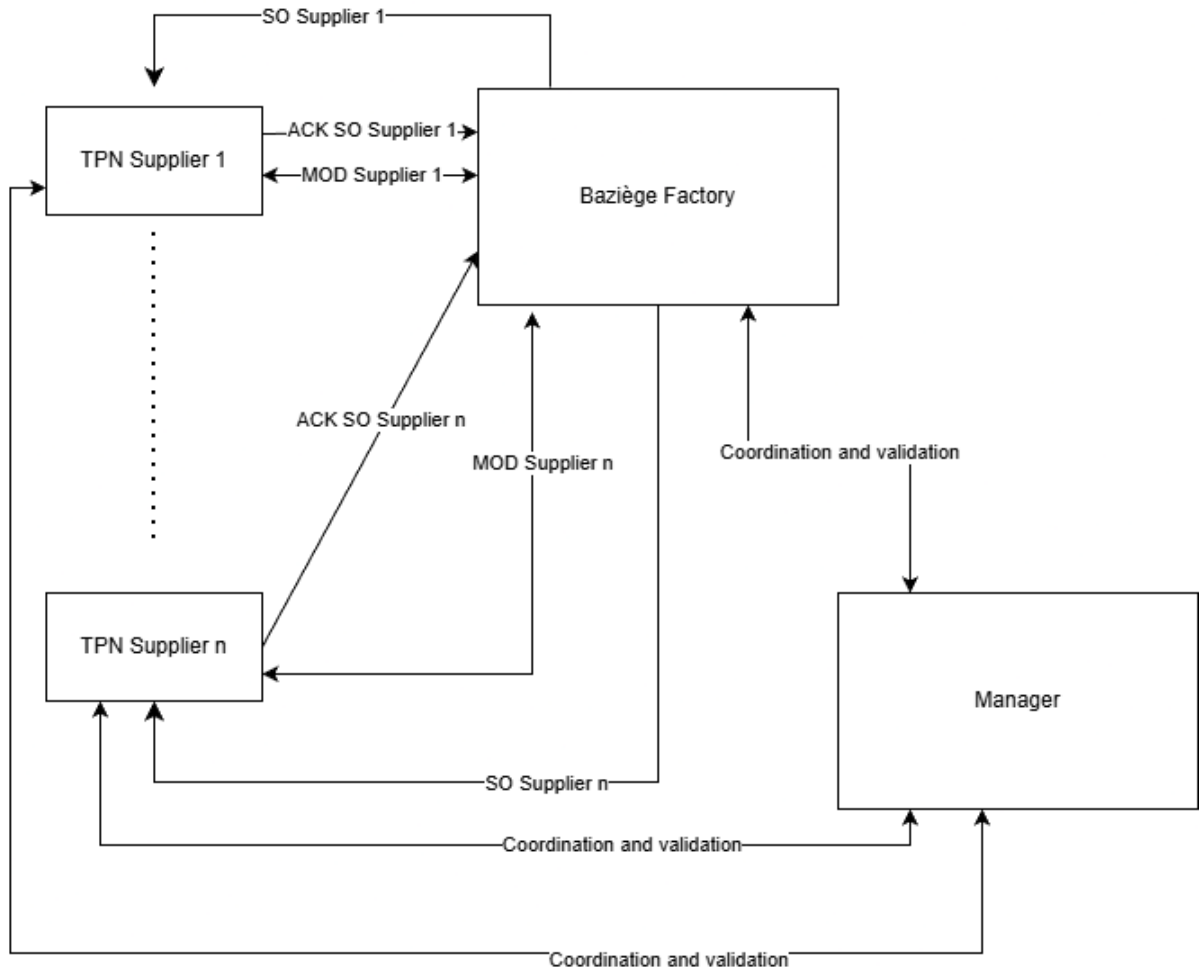}
    \caption{Global view of our supply chain system}
    \label{fig:GlobalView}
\end{figure}

On this model (Figure~\ref{fig:GlobalView}) we have several subsystems:

\begin{itemize}
    \item Suppliers which are representing the different subcontractors manufacturing the different pieces of our product.
    \item The main factory which handles the supply order and the modification demands.
    \item The manager which coordinates and validates the different demands.
\end{itemize}

All of them have interactions which will be represented as labelled transitions to be synchronised in our PTPN.

\begin{itemize}
    \item SO: Supply Order, the action of purchasing a product.
    \item ACK: Acknowledgement.
    \item MOD: Demand of modification for the piece in case of slight difference between the order and the produce piece. The produce is still up to the norm but not to the order.
    \item Coordination and Validation: The Manager will be in the process of validation of pieces, with or without modification.
\end{itemize}

We present the labelled transitions of the TPN such that each label indicates the source and destination in the form of $type\_source\_destination$. As an example, $SO\_BAZ\_Si$ indicates a supply order (SO) from the main factory (BAZ) to the Supplier i (Si). The only exception is for the acknowledgment (ACK), for which we use a distinct labelling to emphasize the acknowledgment more prominently.

In our supply-chain model, each actor is associated with an operation characterised by a bounded execution time, expressed through best-case and worst-case durations. The global supply-chain lead time is then obtained by aggregating the worst-case durations along the execution path from raw material suppliers to final delivery. This corresponds to a pessimistic timing analysis, analogous to worst-case critical-path evaluation in PERT~\cite{malcolm1959pert}.

By considering only worst-case durations, we adopt a conservative modelling approach that is well suited to industrial supply chains, where early completion of an operation does not necessarily reduce the start time of subsequent operations due to synchronisation constraints, inventory policies, or organizational delays. This abstraction is consistent with both PERT-based reasoning and Time Petri Net semantics, and allows us to reason safely about global lead times in a distributed manufacturing context. From this perspective, the proposed TPN-based model can be seen as a generalisation of PERT networks, enriched with concurrency, synchronisation, and formal execution semantics.

The timing intervals associated with transitions in our Time Petri Net model are intended to represent typical processing, transportation, and decision delays observed in supply chain operations, rather than precise measurements of a specific industrial system. This modelling choice is consistent with existing work on Timed Petri Net–based supply chain analysis, where time intervals are used to capture lead times, processing durations, and coordination delays between actors~\cite{Mazzuto15082012}.

In addition, the selection of timing intervals in our model is informed by practical industrial experience, as one of the authors is a supply chain engineer involved in operational planning and coordination activities. The chosen intervals therefore reflect representative values commonly encountered in practice and are used to support the analysis of temporal feasibility, synchronisation constraints, and bottleneck effects, rather than to reproduce a specific operational dataset. Now that we have presented our system as a general concept, we model the different subsystems as TPN representing their timing behaviour. Transitions are represented either as labelled (\textbf{with their label in bold}) or unlabelled (without a bold label). The absence of a timing constraint indicates that the transition has a timing interval of $[0,\infty]$.

\subsection{Supplier}

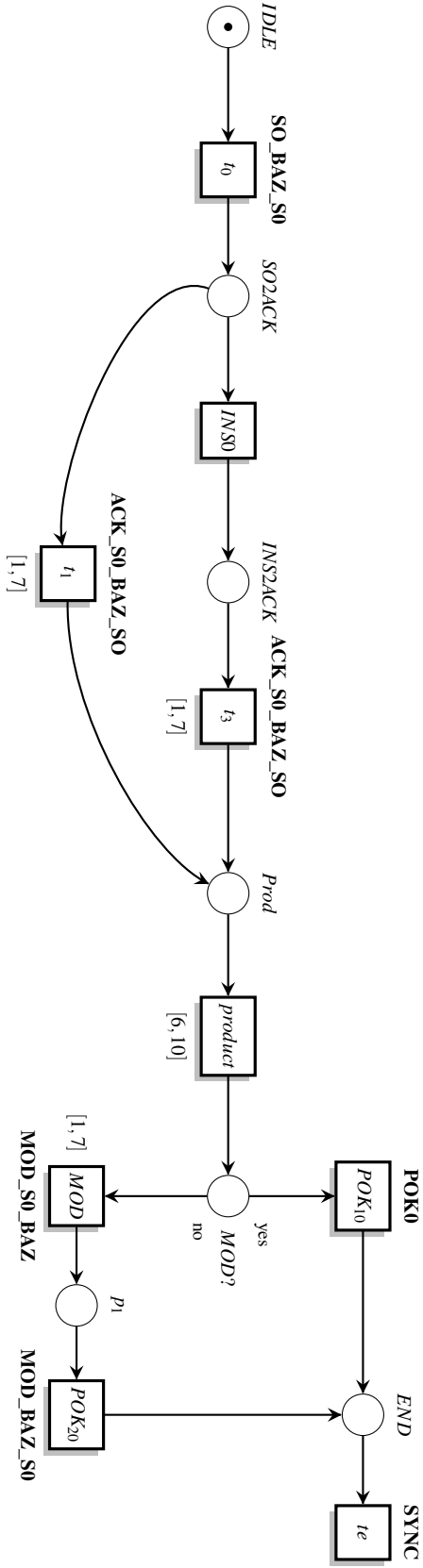
\begin{sidewaysfigure}[p]
\centering
\def\scale{0.675}
\def\scalenodes{1.0}
\begin{tikzpicture}[glob-options]
\node[trans, label=-90:{\large \bf MOD\_S0\_BAZ}, label=180:{\large $[1,7]$}](t0) at (725.0,220.0) {\large $MOD$};
\node[place, label=90:{\large $p_{1}$}](p1) at (790.0, 220.0) {};
\node[trans, label=90:{\large \bf SO\_BAZ\_S0}](t2) at (115.0,130.0) {\large $t_{0}$};
\node[place, label=90:{\large $SO2ACK$}](p3) at (190.0, 130.0) {};
\node[place, label=90:{\large $IDLE$}](p4) at (30.0, 130.0) {};
\node[token] at (p4) {};
\node[trans](t5) at (270.0,130.0) {\large $INS0$};
\node[place, label=90:{\large $INS2ACK$}](p6) at (360.0, 130.0) {};
\node[trans, label=90:{\large \bf ACK\_S0\_BAZ\_SO}, label=-90:{\large $[1,7]$}](t7) at (440.0,130.0) {\large $t_{3}$};
\node[place, label=90:{\large $Prod$}](p8) at (545.0, 130.0) {};
\node[trans, label=90:{\large \bf ACK\_S0\_BAZ\_SO}, label=-90:{\large $[1,7]$}](t9) at (355.0,225.0) {\large $t_{1}$};
\node[trans, label=-90:{\large $[6,10]$}](t10) at (630.0,130.0) {\large $product$};
\node[place, label=0:{\large $MOD?$}, label=45:{yes}, label=-45:{no}](p11) at (725.0, 130.0) {};
\node[trans, label=90:{\large \bf POK0}](t12) at (725.0,50.0) {\large $POK_{10}$};
\node[place, label=90:{\large $END$}](p13) at (855.0, 50.0) {};
\node[trans, label=-90:{\large \bf MOD\_BAZ\_S0}](t14) at (855.0,220.0) {\large $POK_{20}$};
\node[trans, label=90:{\large \bf SYNC}](t15) at (925.0,50.0) {\large $te$};
\draw[arc](p6) -- (t7);
\draw[arc](t5) -- (p6);
\draw[arc](t14) -- (p13);
\draw[arc](p1) -- (t14);
\draw[arc](t0) -- (p1);
\draw[arc](p13) -- (t15);
\draw[arc](t12) -- (p13);
\draw[arc](p11) -- (t12);
\draw[arc](p11) -- (t0);
\draw[arc](p4) -- (t2);
\draw[arc](t2) -- (p3);
\draw[arc](p3) .. controls   +(249.0:61.1pt) and +(191.0:59.0pt)   .. (t9);
\draw[arc](p3) -- (t5);
\draw[arc](t7) -- (p8);
\draw[arc](p8) -- (t10);
\draw[arc](t10) -- (p11);
\draw[arc](t9) .. controls   +(359.0:83.0pt) and +(244.0:52.3pt)   .. (p8);
\end{tikzpicture}\\
\textbf{\small{INS : Inspection \ MOD : Modification}}
\caption{TPN of a Supply Chain for Supplier 0}\label{net:SupplierEx}
\end{sidewaysfigure}

The first TPN to produce was the supplier (see Figure~\ref{net:SupplierEx}). The supplier needs a supply order before becoming active, this order came from the main factory. The goal of the supplier is to produce a piece for a final product which would be produced in the main factory.
Once the order is here, the supplier can decide to proceed to $ACK\_S0\_BAZ\_SO$ to acknowledge the order and decide to do an inspection of the ordered product first via $INS$. This can take from 1 to 7 days.

Such an inspection is to ensure that actual manufacturing parameters could be used, or need to be updated to the latest industrial standard. These inspections are done by the supplier alone. Once the inspection is done and the ACK is sent, the supplier is producing the demanded supply, this takes time, from 6 to 10 days in our example. When the product is ready, if everything is up to the requirement, the supplier can send it via the $POK_{10}$ transition, which is validated by an available manager. Otherwise, the supplier asks for a modification via $MOD\_S0\_BAZ$ to ask for a modification of the contract. We do not account for the event of a failure of production here or the event of an unvalidated modification (which could lead to timeout). The focus is put on timing constraints studies. The manager is synchronised with this demand of modification and he is the one to accept it. 

The last transition of the Supplier is $SYNC$ which synchronise this Supplier with other TPN to conclude on the success of the whole system.

This supplier can be adapted to $n$ others supplier by changing $0$ for another number.


\subsection{Manager}

The main goal of the manager is to validate one piece or allow for modification of the order. The supply chain cannot proceed without the explicit validation of a manager. 
It has two variables:

\begin{itemize}
    \item $x$: the number of managers which are available in the $IDLE$ place.
    \item $y$: the upper bound of the modification of the supply order. This variable represents the maximum amount of time a manager could spend granting a modification which can be a bottleneck in the supply chain system.
\end{itemize}

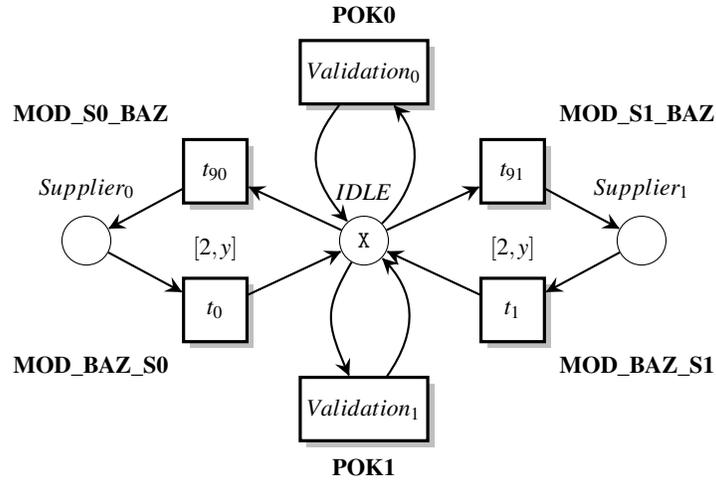
\begin{figure}[H]
\centering
\def\scale{0.75}
\def\scalenodes{1.0}
\begin{tikzpicture}[glob-options]
\node[trans, label=-135:{\large \bf MOD\_BAZ\_S0}, label=90:{\large $[2,y]$}](t18) at (105.0,170.0) {\large $t_{0}$};
\node[trans, label=135:{\large \bf MOD\_S0\_BAZ}](t19) at (105.0,100.0) {\large $t_{90}$};
\node[trans, label=90:{\large \bf POK0}](t20) at (180.0,50.0) {\large $Validation_{0}$};
\node[trans, label=-45:{\large \bf MOD\_BAZ\_S1}, label=90:{\large $[2,y]$}](t21) at (255.0,170.0) {\large $t_{1}$};
\node[trans, label=45:{\large \bf MOD\_S1\_BAZ}](t22) at (255.0,100.0) {\large $t_{91}$};
\node[trans, label=-90:{\large \bf POK1}](t23) at (180.0,220.0) {\large $Validation_{1}$};
\node[place, label=90:{\large $Supplier_{0}$}](p24) at (40.0, 135.0) {};
\node[place, label=90:{\large $IDLE$}](p25) at (180.0, 135.0) {\large \tt X};
\node[place, label=90:{\large $Supplier_{1}$}](p26) at (320.0, 135.0) {};
\draw[arc](p24) -- (t18);
\draw[arc](t18) -- (p25);
\draw[arc](p25) -- (t19);
\draw[arc](t19) -- (p24);
\draw[arc](p25) .. controls   +(43.8:33.2pt) and +(312.0:51.0pt)   .. (t20);
\draw[arc](t20) .. controls   +(233.0:42.8pt) and +(131.0:48.9pt)   .. (p25);
\draw[arc](p26) -- (t21);
\draw[arc](t21) -- (p25);
\draw[arc](p25) -- (t22);
\draw[arc](t22) -- (p26);
\draw[arc](p25) .. controls   +(240.0:40.3pt) and +(114.0:48.3pt)   .. (t23);
\draw[arc](t23) .. controls   +(55.0:48.8pt) and +(316.0:36.1pt)   .. (p25);
\end{tikzpicture}
\caption{TPN of X Managers for two supply chain}\label{net:Manager}
\end{figure}

The manager presented in Figure~\ref{net:Manager} is for two suppliers (0 and 1). To add another supplier, there is a copy of the transitions and places for this new process. We focus on Supplier 0 in this case.

The transition $Validation_0$ is synchronised on the label $POK0$ which validates a piece produced by a supplier. A manager can also be unavailable (which could lead to a \emph{timelock}, see Section~\ref{sec:sAa}). The transitions $t_{90}$ is synchronised on $MOD\_S0\_BAZ$ and is a demand of modification from a supplier. The manager then has to go to this supplier to check the modification and grant it with the transition $t_{0}$ via the synchronised label $MOD\_BAZ\_S0$.

The manager has a key role in allowing the flow of pieces from supplier to factory. 

\subsection{Factory}

\begin{figure}[H]
\centering
\def\scale{0.60}
\def\scalenodes{1.0}
\begin{tikzpicture}[glob-options]
\node[trans, label=90:{\large \bf SO\_BAZ\_S0}, label=-90:{\large $[0,1]$}](t0) at (135.0,50.0) {\large $t_{0}$};
\node[place, label=90:{\large $SO2ACK$}](p1) at (230.0, 50.0) {};
\node[trans, label=90:{\large \bf ACK\_S0\_BAZ\_SO}](t2) at (320.0,50.0) {\large $t_{1}$};
\node[place, label=90:{\large $MOD0?$}](p3) at (420.0, 50.0) {};
\node[trans, label=180:{\large \bf MOD\_S0\_BAZ}](t4) at (420.0,130.0) {\large $MOD_{0}$};
\node[place, label=-90:{\large $Modification_{0}$}](p5) at (515.0, 130.0) {};
\node[trans, label=-90:{\large \bf POK0}](t6) at (515.0,50.0) {\large $POK_{10}$};
\node[place, label=90:{\large $p_{3}$}](p7) at (620.0, 50.0) {};
\node[trans, label=0:{\large \bf MOD\_BAZ\_S0}](t8) at (620.0,130.0) {\large $POK_{20}$};
\node[trans, label=90:{\large \bf SYNC}, label=-90:{\large $[0,0]$}](t9) at (720.0,50.0) {\large $te$};
\node[place, label=90:{\large $IDLE_{0}$}](p10) at (30.0, 50.0) {};
\node[token] at (p10) {};
\node[trans, label=90:{\large \bf SO\_BAZ\_S1}, label=-90:{\large $[0,1]$}](t11) at (135.0,195.0) {\large $t_{10}$};
\node[place, label=90:{\large $SO2ACK_{1}$}](p12) at (230.0, 195.0) {};
\node[trans, label=90:{\large \bf ACK\_S1\_BAZ\_SO}](t13) at (320.0,195.0) {\large $t_{11}$};
\node[place, label=90:{\large $MOD1?$}](p14) at (420.0, 195.0) {};
\node[trans, label=180:{\large \bf MOD\_S1\_BAZ}](t15) at (420.0,275.0) {\large $MOD_{1}$};
\node[place, label=-90:{\large $Modification_{1}$}](p16) at (515.0, 275.0) {};
\node[trans, label=-90:{\large \bf POK1}](t17) at (515.0,195.0) {\large $POK_{11}$};
\node[place, label=90:{\large $p_{4}$}](p18) at (620.0, 195.0) {};
\node[trans, label=0:{\large \bf MOD\_BAZ\_S1}](t19) at (620.0,275.0) {\large $POK_{21}$};
\node[place, label=90:{\large $IDLE_{1}$}](p20) at (30.0, 195.0) {};
\node[token] at (p20) {};
\draw[arc](p18) .. controls   +(0.494:116.0pt) and +(350:100pt)   .. (t9);
\draw[arc](t19) -- (p18);
\draw[arc](p16) -- (t19);
\draw[arc](t17) -- (p18);
\draw[arc](p14) -- (t17);
\draw[arc](t15) -- (p16);
\draw[arc](p14) -- (t15);
\draw[arc](t13) -- (p14);
\draw[arc](p12) -- (t13);
\draw[arc](t11) -- (p12);
\draw[arc](p20) -- (t11);
\draw[arc](p10) -- (t0);
\draw[arc](t0) -- (p1);
\draw[arc](p1) -- (t2);
\draw[arc](t2) -- (p3);
\draw[arc](p3) -- (t4);
\draw[arc](t4) -- (p5);
\draw[arc](p3) -- (t6);
\draw[arc](t6) -- (p7);
\draw[arc](p5) -- (t8);
\draw[arc](t8) -- (p7);
\draw[arc](p7) -- (t9);
\end{tikzpicture}
\caption{TPN of the Factory for two suppliers}\label{net:BAZEx}
\end{figure}
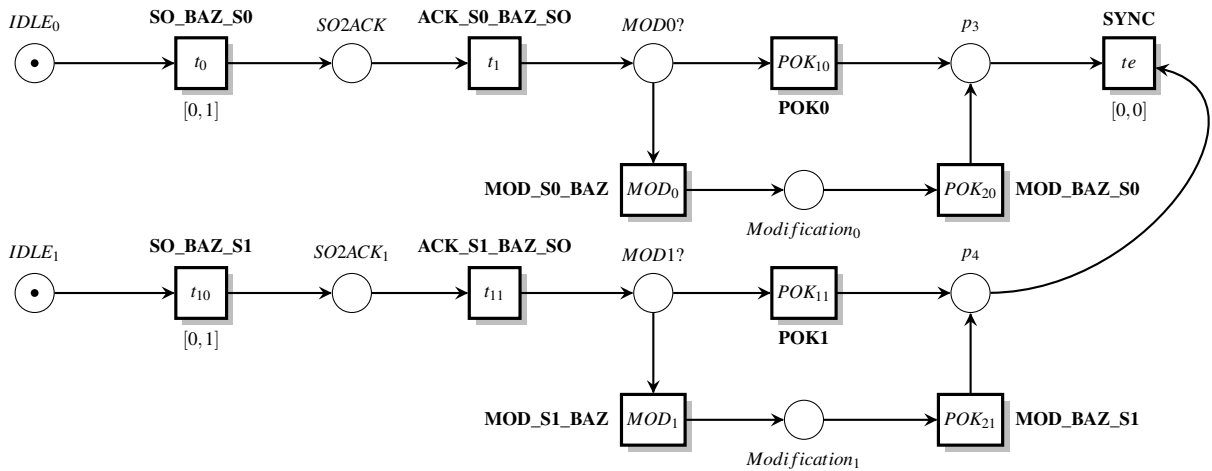

Our factory (called $BAZ$ in the TPN) is the entity asking for supply. We present in Figure~\ref{net:BAZEx} the Factory behaviour for two suppliers. To handle more suppliers, we simply add one more path with a new identification ($2$ for Supplier 2 for example).

A factory process begins by asking for a Supply Order (SO) via the first transition and the label $SO\_BAZ\_S0$. This step can take up to a day. At this point, the factory receives most of its information from the supplier (acknowledging the supply order) and the manager (acceptation modification via $MOD\_BAZ\_Si$) net, following their processes until the last transition. All of the factory paths are synchronised in the $t_e$ transition which is active when all of the suppliers are done handling the supply orders. This transition is urgent and it is represented as a $[0,0]$ timing constraint. 

A key component of the Factory is that it almost orders the supply simultaneously. 

The goal of this paper was also to model-check some properties on our synchronised product. In the next section we are adding an ending net, called \emph{end-of-line} to synchronise the overall system and check the viability of our supply chain model.

\subsection{End-of-line}

The end-of-line net is synchronised at the end of the supply order via the $SYNC$ label. The single token in the Waiting place allows the $timeout$ transition to have time elapse. The 210 days correspond to the accepted global supply-chain lead time by the factory. Once another token is put into the Waiting place, the $t_0$ transition became sensibilised. Since it is an urgent transition it must fire immediately. The $success$ transition is checked to validate the viability of the supply chain.

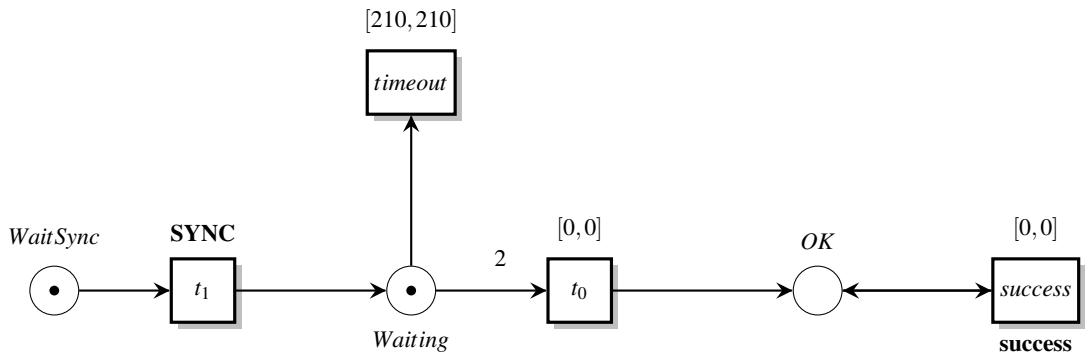
\begin{figure}[H]
\centering
\def\scale{0.75}
\def\scalenodes{1.0}
\begin{tikzpicture}[glob-options]
\node[place, label=90:{\large $OK$}](p0) at (425.0, 155.0) {};
\node[trans, label=-90:{\large \bf success}, label=90:{\large $[0,0]$}](t1) at (535.0,155.0) {\large $success$};
\node[place, label=-90:{\large $Waiting$}](p2) at (220.0, 155.0) {};
\node[token] at (p2) {};
\node[trans, label=90:{\large \bf SYNC}](t3) at (115.0,155.0) {\large $t_{1}$};
\node[trans, label=90:{\large $[0,0]$}](t4) at (305.0,155.0) {\large $t_{0}$};
\node[trans, label=90:{\large $[210,210]$}](t5) at (220.0,50.0) {\large $timeout$};
\node[place, label=90:{\large $WaitSync$}](p6) at (40.0, 155.0) {};
\node[token] at (p6) {};
\draw[arc](t1) -- (p0);
\draw[arc](p0) -- (t1);
\draw[arc](p2) -- (t4);
\path(t4)+(180.0:40.0pt) node [label=90:\large 2] {};
\draw[arc](p2) -- (t5);
\draw[arc](t4) -- (p0);
\draw[arc](t3) -- (p2);
\draw[arc](p6) -- (t3);
\end{tikzpicture}
\caption{TPN end-of-line}\label{net:end}
\end{figure}

This net is more independent from the system since the only synchronisation is on the first transition. Now that we have introduced our net, we produce a full model via a PTPN product.

\section{Experimental Results}
\label{sec:sAa}

Our experiment follows several steps:
\begin{itemize}
    \item First, a model is generated from the TPN representing our supply chain.
    \begin{itemize}
        \item The firing intervals of the manager transitions (with $[2,y]$) are considered as variables.
        \item The number of managers and suppliers are also considered as a variables.
    \end{itemize}
    \item Second, we use a model checker to determine the feasibility of the resulting supply chain model.
\end{itemize}

\begin{figure}[H]
    \centering
    \includegraphics[width=0.75\linewidth]{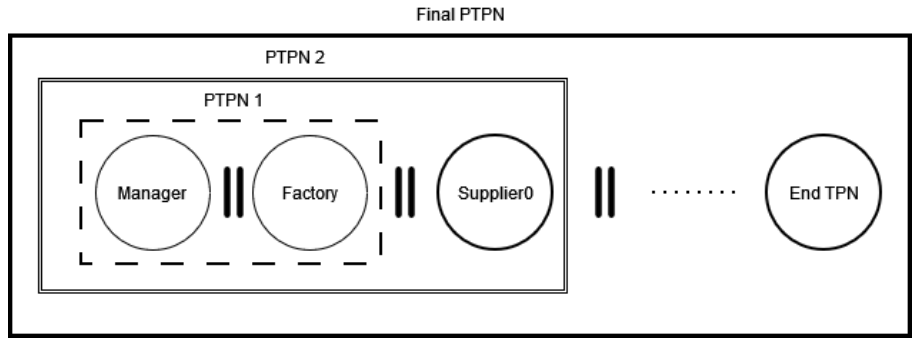}
    \caption{Process of Synchronisation of the TPN}
    \label{fig:Synchro}
\end{figure}

We synchronise our models following the process described in Figure~\ref{fig:Synchro}. Since PTPNs are composable, the synchronisation is performed incrementally, net by net. First, the manager and the factory are synchronised, then each supplier is added one by one, and finally the TPN representing the end-of-line is incorporated.

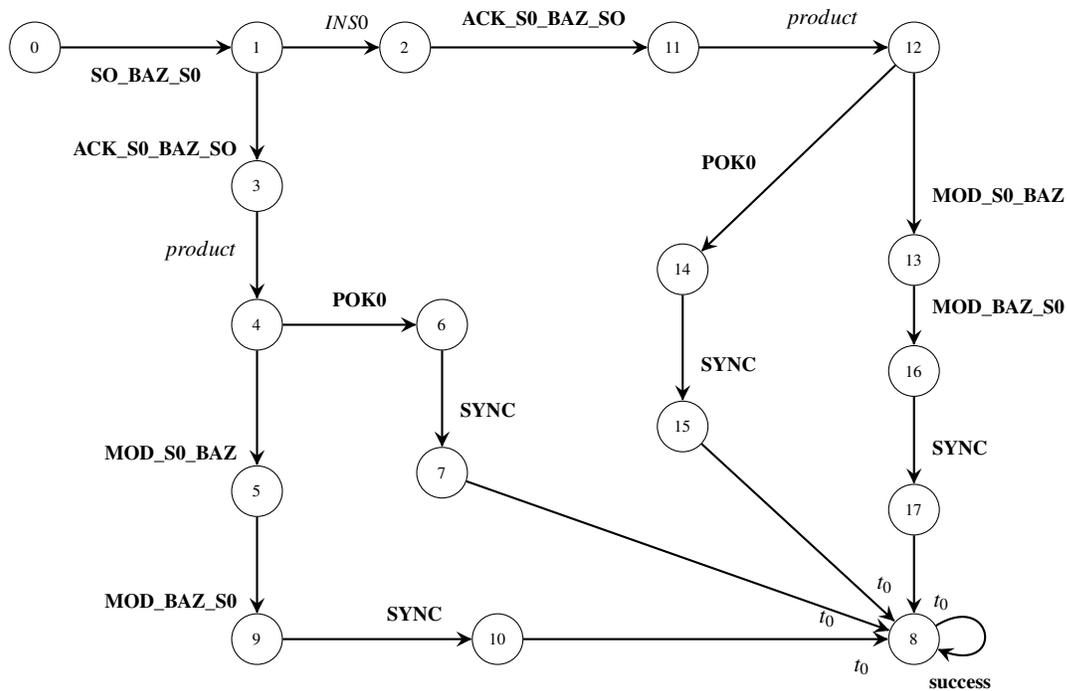
\begin{figure}[H]
\centering
\hspace*{-1cm}
\def\scale{0.7}
\def\scalenodes{0.75}
\begin{tikzpicture}[glob-options]
\node[state](s0) at (455.0, 30.0) {\large $_{11}$};
\node[state](s1) at (585.0, 280.0) {\large $_{17}$};
\node[state](s2) at (585.0, 145.0) {\large $_{13}$};
\node[state](s3) at (230.0, 105.0) {\large $_{3}$};
\node[state](s4) at (585.0, 30.0) {\large $_{12}$};
\node[state](s5) at (310.0, 30.0) {\large $_{2}$};
\node[state](s6) at (230.0, 30.0) {\large $_{1}$};
\node[state](s7) at (110.0, 30.0) {\large $_{0}$};
\node[state](s8) at (460.0, 150.0) {\large $_{14}$};
\node[state](s9) at (460.0, 235.0) {\large $_{15}$};
\node[state](s10) at (585.0, 205.0) {\large $_{16}$};
\node[state](s11) at (230.0, 180.0) {\large $_{4}$};
\node[state](s12) at (330.0, 180.0) {\large $_{6}$};
\node[state](s13) at (330.0, 260.0) {\large $_{7}$};
\node[state](s14) at (230.0, 270.0) {\large $_{5}$};
\node[state](s15) at (360.0, 350.0) {\large $_{10}$};
\node[state](s16) at (230.0, 350.0) {\large $_{9}$};
\node[state](s17) at (585.0, 350.0) {\large $_{8}$};
\draw[arc](s1) -- (s17);
\path(s17)+(90.0:20.0pt) node [label=0:{\large $t_0$}] {};
\draw[arc](s10) -- (s1);
\path(s1)+(90.0:20.0pt) node [label=45:{\textbf{SYNC}}] {};
\draw[arc](s9) -- (s17);
\path(s17)+(138.0:21.2pt) node [label=90:{\large $t_0$}] {};
\draw[arc](s8) -- (s9);
\path(s9)+(90.0:20.0pt) node [label=45:{\textbf{SYNC}}] {};
\draw[arc](s2) -- (s10);
\path(s10)+(90.0:20.0pt) node [label=45:{\textbf{MOD\_BAZ\_S0}}] {};
\draw[arc](s4) -- (s8);
\path(s8)+(60:50pt) node [label=90:{\textbf{POK0}}] {};
\draw[arc](s4) -- (s2);
\path(s2)+(90.0:20.0pt) node [label=45:{\textbf{MOD\_S0\_BAZ}}] {};
\draw[arc](s0) -- (s4);
\path(s4)+(180.0:20.0pt) node [label=135:{\large $product$}] {};
\draw[arc](s15) -- (s17);
\path(s17)+(180.0:28.1pt) node [label=-90:{\large $t_0$}] {};
\draw[arc](s16) -- (s15);
\path(s15)+(180.0:20.0pt) node [label=135:{\textbf{SYNC}}] {};
\draw[arc](s17) .. controls   +(31.5:51.6pt) and +(337.0:54.2pt)   .. (s17);
\path(s17)+(337.0:27.1pt) node [label=-90:{\textbf{success}}] {};
\draw[arc](s13) -- (s17);
\path(s17)+(161.0:33.8pt) node [label=180:{\large $t_0$}] {};
\draw[arc](s12) -- (s13);
\path(s13)+(90.0:20.0pt) node [label=45:{\textbf{SYNC}}] {};
\draw[arc](s14) -- (s16);
\path(s16)+(90.0:20.0pt) node [label=180:{\textbf{MOD\_BAZ\_S0}}] {};
\draw[arc](s11) -- (s12);
\path(s12)+(180.0:20.0pt) node [label=135:{\textbf{POK0}}] {};
\draw[arc](s11) -- (s14);
\path(s14)+(90.0:20.0pt) node [label=180:{\textbf{MOD\_S0\_BAZ}}] {};
\draw[arc](s3) -- (s11);
\path(s11)+(90.0:40.0pt) node [label=180:{\large $product$}] {};
\draw[arc](s5) -- (s0);
\path(s0)+(180.0:70.0pt) node [label=90:{\textbf{ACK\_S0\_BAZ\_SO}}] {};
\draw[arc](s6) -- (s3);
\path(s3)+(90.0:20.0pt) node [label=180:{\textbf{ACK\_S0\_BAZ\_SO}}] {};
\draw[arc](s6) -- (s5);
\path(s5)+(180.0:10.0pt) node [label=135:{$INS0$}] {};
\draw[arc](s7) -- (s6);
\path(s6)+(180.0:60.0pt) node [label=-90:{\textbf{SO\_BAZ\_S0}}] {};
\end{tikzpicture}
    \caption{SCG of the product}
    \label{fig:Kripke}
\end{figure}

For a single supplier, a single manager, and a timing interval of $[2,6]$ on the manager transition $t_0$, we obtain the SCG of the product, shown in Figure~\ref{fig:Kripke}. Each state of this SCG encompasses both the marking and the firing domain of the underlying PTPN. This SCG represents a viable supply chain that always leads to successful delivery, as no deadlocks occur. Deadlocks would result from violations of timing constraints or insufficient managerial resources.

To support reproducibility and reuse, the benchmark generator and the end-of-line Petri net are publicly available online\footnote{\url{https://github.com/Darkelubat/SupplyChain}}\footnote{\url{https://zenodo.org/records/18925093}}. 
Our benchmark was executed on an MSI GS70 equipped with an Intel Core i5 processor. 
We use TWINA to synchronise our TPNs using the following command:
\begin{lstlisting}
	twina -aut fuse.tpn
\end{lstlisting}

TWINA processes the SCG defined in the \emph{fuse.tpn} file. If two transitions have the same label and name, TWINA adds a suffix to distinguish them (e.g., $\{t_0, t_0\}$ becomes $\{t_{0}.1, t_{0}.2\}$).

We considered several configurations ranging from 1 to 3 suppliers. The synchronisation process may produce a \emph{timelock}. In our setting, such a situation is not caused by a timeout but rather by an insufficient number of managers.

One of the main variables in our model is the timing constraint applied to the Manager model. We vary the upper bound of the firing interval to determine when the model can no longer produce a successful execution. Failures may occur either because of a timeout (via the \emph{timeout} transitions in the net) or because of a \emph{timelock}, indicating that the timing constraints are too restrictive for the supply chain given the current number of managers.

To detect these situations, we use the following formulas in \emph{selt}:
\begin{lstlisting}
	selt PTPN.ktz -f "<> [] {success}" -v
	selt PTPN.ktz -f "-dead \/ <>{timeout}" -v
\end{lstlisting}

A single occurrence of a timeout is sufficient to conclude that it may occur, thereby rendering the supply chain non-viable. 
Our \emph{selt} formulas can be expressed both textually and in classical LTL notation as follows:

\medskip
\noindent
\textbf{1. Success property:} Does the success transition eventually hold forever? In LTL:
\[
\lozenge \square \text{success}
\]

\noindent
\textbf{2. Deadlock/Timeout property:} Is there a deadlock or does a timeout eventually occur? In LTL:
\[
\text{dead} \lor \lozenge \text{timeout}
\]

Table~\ref{fig:SyncRes} summarises the time required to compute the product. 
For each configuration, the table reports the output of the TWINA command, including the processing time, the number of classes, markings, firing domains, and transitions composing the resulting PTPN model.

\begin{table}[h]
    \centering
    \begin{tabular}{|p{4cm}|p{2cm}|p{2cm}|p{2cm}|p{2cm}|}
    \hline
     Supplier & 1 & 2 & 3 & 4 \\
     \hline     \hline
     Time  & 0.006s & 0.044s & 5.7s & 8m12s\\
     \hline
     Classes & 18 & 622 & 42136 &  2760432\\
     \hline
     Marking & 9 & 44 & 239 & 1302\\
     \hline
     Firing Domains & 18 & 588 & 37585 & 2320772\\
     \hline
     Transitions & 21 & 1001 & 98769 &  8385033\\
     \hline     
    \end{tabular}
    \caption{Processing of the final PTPN depending on the number of Supplier}    
    \label{fig:SyncRes}
\end{table}

While processing a synchronised supply chain system, two issues can arise:

\begin{itemize}
    \item The number of suppliers for a single manager will lead to management issues, so we scale it down for further experimentations and also test it with more managers.
    \item Combinatorial explosion is still an issue when synchronising several TPNs.
\end{itemize}

We analysed our PTPN model using \emph{PTPN.ktz}. If one formula evaluate to a failure, we follow the counterexample provided by \emph{selt} to identify a \emph{timelock}. The results are summarised in Table~\ref{fig:Table}, where $TimeOut$ denotes a timeout of the supply order and $Timelock$ indicates a deadlock arising from insufficient managerial resources.

\begin{table}[h]
    \centering
    \begin{tabular}{|p{2cm}||p{2cm}|p{2cm}|p{2cm}|p{2cm}|p{2cm}|}
     \hline
     \multicolumn{6}{|c|}{Supply chain feasibility} \\
     \hline
     Supplier & 1 & 2 & 2 & 3 & 3\\
     \hline
     Manager & 1 & 1 & 2 & 2 & 3\\
     \hline
     \multicolumn{6}{|l|}{$[2,y]$} \\
     \hline
     $[2,6]$   & Success & TimeLock & Success & TimeLock & Success \\
     $[2,15]$   & Success & TimeLock & Success & TimeLock & Success\\
     $[2,50]$ &   Success  & TimeLock   & Success & TimeLock & Success\\
     $[2,60]$ &   Success  & TimeLock   & Success & TimeLock & Success\\
     $[2,175]$ &  Success  & TimeLock   & Success & TimeLock & Success\\
     $[2,180]$ &  TimeOut  & TimeOut   & TimeOut & TimeOut & TimeOut\\
     \hline
    \end{tabular}
    \caption{Supply Chain feasibility depending on the number of Suppliers, Managers and timing constraints}
    \label{fig:Table}
\end{table}

The main issue is that the manager is unable to handle more than one supplier at a time, which leads to a \emph{timelock}. 
As discussed in Section~\ref{sec:sc}, introducing slight timing differences in supply orders could allow a single manager to handle several suppliers more efficiently. Staggering supply orders reduces simultaneous demands on the manager’s modification transitions, thereby avoiding incompatible timing constraints that would otherwise lead to timelocks.

To test this hypothesis, we modified the timing constraints on the first transitions of the factory as follows:
\begin{itemize}
    \item $t_0$ remains $[0,1]$,
    \item $t_{10}$ is set to $[50,100]$.
\end{itemize}

The resulting measurements are reported in Table~\ref{fig:TableMod}.

\begin{table}[h]
    \centering
    \begin{tabular}{|p{2cm}||p{2cm}|p{2cm}|p{2cm}|}
     \hline
     \multicolumn{4}{|c|}{Supply chain feasibility} \\
     \hline
     Supplier & 1 & 2 & 2 \\
     \hline
     Manager & 1 & 1 & 2\\
     \hline
     \multicolumn{4}{|l|}{$[2,y]$} \\
     \hline
     $[2,6]$   & Success & Success & Success \\
     $[2,15]$   & Success & Success & Success \\
     $[2,50]$ &   Success  & Success   & Success\\
     $[2,60]$ &   Success  & Success   & Success\\
     $[2,175]$ &  Success  & TimeOut   & TimeOut\\
     $[2,180]$ &  TimeOut  & TimeOut   & TimeOut\\
     \hline
    \end{tabular}
    \caption{Supply Chain feasibility depending on the number of Suppliers, Managers, and timing constraints with the New Staggered Supply Order}
    \label{fig:TableMod}
\end{table}

As we can see, a better handling of supply orders clearly improves the manager's ability to manage multiple suppliers. In some configurations, however, \emph{timeouts} now occur because certain suppliers are delayed enough to affect the overall system behaviour. The factory must therefore balance the ordering of supplies with the time required to perform the necessary modifications.

\section{Conclusion}
\label{sec:conclusion}

In this paper, we proposed a modular approach for modelling and analysing supply chains based on Product Time Petri Nets. Each subsystem of the supply chain composed of suppliers, factory, and supply manager was modelled independently as a Time Petri Net, and their interactions were captured through synchronised transition labels. This compositional approach allows complex supply chain behaviours to be constructed incrementally while preserving precise timing semantics.

A key aspect of our model is the explicit representation of the Supply Chain and supply Manager as a critical shared resource. This makes it possible to study not only delay propagation across suppliers, but also the impact of timing constraints and managerial availability on the overall feasibility of the supply chain. Using the TINA toolbox and LTL-based verification with \emph{selt}, we showed how timing constraints and resource allocation jointly influence system outcomes, leading either to successful completion, timeout failures, or timelocks induced by incompatible timing constraints.

Our experimental results highlight several important insights. First, increasing the number of suppliers without increasing managerial capacity quickly leads to infeasible configurations, even in the absence of explicit timeouts. Second, the timing of supply orders plays a crucial role: staggering orders can significantly improve feasibility by reducing contention on shared managerial resources. These observations illustrate how PTPN-based models can support what-if analyses for supply chain design and decision-making.

As expected, the synchronised composition of multiple timed components leads to a rapid growth of the state space, which currently limits exhaustive analysis to a small number of suppliers. However, this limitation does not affect the scalability of the modelling approach itself, which remains modular and reusable. Rather, it highlights the need for future work on compositional analysis techniques, partial-order reductions, or abstraction methods tailored to synchronised Time Petri Nets.

Future work will explore several directions. First, we plan to extend the model to represent the flow of multiple pieces, rather than focusing solely on timing constraints. Second, additional supplier behaviours, with failures, rework loops, storage, alternative validation paths could be incorporated to increase realism. Finally, PTPN-based supply chain models provide a promising benchmark to investigate scalable verification techniques for synchronised timed systems.

\subsection*{Acknowledgements}

We are grateful to Uli FAHRENBERG for discussing the modelling of Supply Chain, to Adrien VINEL for his insight on manufacturing, to Bernard BERTHOMIEU and Silvano DAL ZILIO for their help with the TINA toolbox.

\bibliographystyle{eptcs}
\bibliography{generic}

\newcommand{\Afirst}[1]{#1} \newcommand{\afirst}[1]{#1}
\begin{thebibliography}{10}
\providecommand{\bibitemdeclare}[2]{}
\providecommand{\surnamestart}{}
\providecommand{\surnameend}{}
\providecommand{\urlprefix}{Available at }
\providecommand{\url}[1]{\texttt{#1}}
\providecommand{\href}[2]{\texttt{#2}}
\providecommand{\urlalt}[2]{\href{#1}{#2}}
\providecommand{\doi}[1]{doi:\urlalt{https://doi.org/#1}{#1}}
\providecommand{\eprint}[1]{arXiv:\urlalt{https://arxiv.org/abs/#1}{#1}}
\providecommand{\bibinfo}[2]{#2}

\bibitemdeclare{inproceedings}{vanderAalst1994ModellingAA}
\bibitem{vanderAalst1994ModellingAA}
\bibinfo{author}{Wil~M.P. \surnamestart van~der Aalst\surnameend}
  (\bibinfo{year}{1994}): \emph{\bibinfo{title}{Modelling and analysing
  workflow using a Petri-net based approach}}.
\newblock \urlprefix\url{https://api.semanticscholar.org/CorpusID:376304}.

\bibitemdeclare{article}{WorkflowAalst}
\bibitem{WorkflowAalst}
\bibinfo{author}{Wil~M.P. \surnamestart van~der Aalst\surnameend}
  (\bibinfo{year}{1998}): \emph{\bibinfo{title}{The Application of Petri Nets
  to Workflow Management}}.
\newblock {\slshape \bibinfo{journal}{Journal of Circuits, Systems, and
  Computers}} \bibinfo{volume}{8}, pp. \bibinfo{pages}{21--66},
  \doi{10.1142/S0218126698000043}.

\bibitemdeclare{inproceedings}{berard2005comparison}
\bibitem{berard2005comparison}
\bibinfo{author}{B{\'e}atrice \surnamestart B{\'e}rard\surnameend},
  \bibinfo{author}{Franck \surnamestart Cassez\surnameend},
  \bibinfo{author}{Serge \surnamestart Haddad\surnameend},
  \bibinfo{author}{Didier \surnamestart Lime\surnameend} \&
  \bibinfo{author}{Olivier~H \surnamestart Roux\surnameend}
  (\bibinfo{year}{2005}): \emph{\bibinfo{title}{Comparison of the
  expressiveness of timed automata and time {Petri} nets}}.
\newblock In: {\slshape \bibinfo{booktitle}{Formal {Modeling} and {Analysis} of
  {Timed} {Systems} (FORMATS)}}, {\slshape \bibinfo{series}{LNCS}}
  \bibinfo{volume}{3829}, \bibinfo{organization}{Springer},
  \doi{10.1007/978-3-540-85778-5_3}.

\bibitemdeclare{article}{BD91}
\bibitem{BD91}
\bibinfo{author}{B.~\surnamestart Berthomieu\surnameend} \&
  \bibinfo{author}{M.~\surnamestart Diaz\surnameend} (\bibinfo{year}{1991}):
  \emph{\bibinfo{title}{Modeling and Verification of Time Dependent Systems
  Using Time {P}etri Nets.}}
\newblock {\slshape \bibinfo{journal}{IEEE Trans. on Software Engineering}}
  \bibinfo{volume}{17}(\bibinfo{number}{3}), \doi{10.1109/32.75415}.

\bibitemdeclare{inproceedings}{berthomieu1983enumerative}
\bibitem{berthomieu1983enumerative}
\bibinfo{author}{B.~\surnamestart Berthomieu\surnameend} \&
  \bibinfo{author}{M.~\surnamestart Menasche\surnameend}
  (\bibinfo{year}{1983}): \emph{\bibinfo{title}{An enumerative approach for
  analyzing time {Petri} nets}}.
\newblock In: {\slshape \bibinfo{booktitle}{Proceedings IFIP}}.

\bibitemdeclare{inproceedings}{BPV06}
\bibitem{BPV06}
\bibinfo{author}{B.~\surnamestart Berthomieu\surnameend},
  \bibinfo{author}{F.~\surnamestart Peres\surnameend} \&
  \bibinfo{author}{F.~\surnamestart Vernadat\surnameend}
  (\bibinfo{year}{2006}): \emph{\bibinfo{title}{Bridging the Gap Between Timed
  Automata and Bounded Time {Petri} Nets}}.
\newblock In: {\slshape \bibinfo{booktitle}{Formal {Modeling} and {Analysis} of
  {Timed} {Systems} (FORMATS)}}, {\slshape \bibinfo{series}{LNCS}}
  \bibinfo{volume}{4202}, \bibinfo{organization}{Springer},
  \doi{10.1007/11867340_7}.

\bibitemdeclare{inproceedings}{BPV07}
\bibitem{BPV07}
\bibinfo{author}{B.~\surnamestart Berthomieu\surnameend},
  \bibinfo{author}{F.~\surnamestart Peres\surnameend} \&
  \bibinfo{author}{F.~\surnamestart Vernadat\surnameend}
  (\bibinfo{year}{2007}): \emph{\bibinfo{title}{Model Checking Bounded
  Prioritized Time {P}etri Nets}}.
\newblock In: {\slshape \bibinfo{booktitle}{5th Int. Symp. on Automated
  Technology for Verification and Analysis}}, \bibinfo{series}{LNCS},
  \bibinfo{organization}{Springer}, \doi{10.1007/978-3-540-75596-8_37}.

\bibitemdeclare{article}{berthomieu2004}
\bibitem{berthomieu2004}
\bibinfo{author}{Bernard \surnamestart Berthomieu\surnameend},
  \bibinfo{author}{P.-O \surnamestart Ribet\surnameend} \&
  \bibinfo{author}{Francois \surnamestart Vernadat\surnameend}
  (\bibinfo{year}{2004}): \emph{\bibinfo{title}{The tool {TINA} -- Construction
  of Abstract State Spaces for {P}etri Nets and Time {P}etri Nets}}.
\newblock {\slshape \bibinfo{journal}{International Journal of Production
  Research}} \bibinfo{volume}{42}(\bibinfo{number}{14}),
  \doi{10.1080/00207540412331312688}.

\bibitemdeclare{article}{cassez2006structural}
\bibitem{cassez2006structural}
\bibinfo{author}{Franck \surnamestart Cassez\surnameend} \&
  \bibinfo{author}{Olivier~H \surnamestart Roux\surnameend}
  (\bibinfo{year}{2006}): \emph{\bibinfo{title}{Structural translation from
  time {Petri} nets to timed automata}}.
\newblock {\slshape \bibinfo{journal}{Journal of Systems and Software}}
  \bibinfo{volume}{79}(\bibinfo{number}{10}), \doi{10.1016/j.jss.2005.12.021}.

\bibitemdeclare{article}{LIU2007761}
\bibitem{LIU2007761}
\bibinfo{author}{Rong \surnamestart Liu\surnameend}, \bibinfo{author}{Akhil
  \surnamestart Kumar\surnameend} \& \bibinfo{author}{Wil \surnamestart {van
  der Aalst}\surnameend} (\bibinfo{year}{2007}): \emph{\bibinfo{title}{A formal
  modeling approach for supply chain event management}}.
\newblock {\slshape \bibinfo{journal}{Decision Support Systems}}
  \bibinfo{volume}{43}(\bibinfo{number}{3}), pp. \bibinfo{pages}{761--778},
  \doi{10.1016/j.dss.2006.12.009}.
\newblock
  \urlprefix\url{https://www.sciencedirect.com/science/article/pii/S0167923606002144}.

\bibitemdeclare{phdthesis}{lubat:tel-03528121}
\bibitem{lubat:tel-03528121}
\bibinfo{author}{Eric \surnamestart Lubat\surnameend} (\bibinfo{year}{2021}):
  \emph{\bibinfo{title}{{Synchronous Product of Time Petri Nets and its
  Applications to Fault-Diagnosis}}}.
\newblock \bibinfo{type}{Theses}, \bibinfo{school}{{INSA de Toulouse}}.
\newblock \urlprefix\url{https://laas.hal.science/tel-03528121}.

\bibitemdeclare{inproceedings}{LubatDBPS19}
\bibitem{LubatDBPS19}
\bibinfo{author}{{\'{E}}ric \surnamestart Lubat\surnameend},
  \bibinfo{author}{Silvano \surnamestart {Dal Zilio}\surnameend},
  \bibinfo{author}{Didier~Le \surnamestart Botlan\surnameend},
  \bibinfo{author}{Yannick \surnamestart Pencol{\'{e}}\surnameend} \&
  \bibinfo{author}{Audine \surnamestart Subias\surnameend}
  (\bibinfo{year}{2019}): \emph{\bibinfo{title}{A State Class Construction for
  Computing the Intersection of Time {Petri} Nets Languages}}.
\newblock In: {\slshape \bibinfo{booktitle}{Formal Modeling and Analysis of
  Timed Systems ({FORMATS})}}, {\slshape \bibinfo{series}{LNCS}}
  \bibinfo{volume}{11750}, \bibinfo{publisher}{Springer},
  \doi{10.1007/978-3-030-29662-9\_5}.

\bibitemdeclare{inproceedings}{lubatCDC2020}
\bibitem{lubatCDC2020}
\bibinfo{author}{{\'E}ric \surnamestart Lubat\surnameend},
  \bibinfo{author}{Silvano \surnamestart Dal~Zilio\surnameend},
  \bibinfo{author}{Didier \surnamestart Le~Botlan\surnameend},
  \bibinfo{author}{Yannick \surnamestart Pencol{\'e}\surnameend} \&
  \bibinfo{author}{Audine \surnamestart Subias\surnameend}
  (\bibinfo{year}{2020}): \emph{\bibinfo{title}{{A New Product Construction for
  the Diagnosability of Patterns in Time Petri Net}}}.
\newblock In: {\slshape \bibinfo{booktitle}{{59th Conference on Decision and
  Control (CDC) 2020}}}, \bibinfo{address}{Jeju Island (virtual conference),
  South Korea}, \doi{10.1109/CDC42340.2020.9303826}.
\newblock \urlprefix\url{https://laas.hal.science/hal-02989834}.

\bibitemdeclare{article}{malcolm1959pert}
\bibitem{malcolm1959pert}
\bibinfo{author}{D.~G. \surnamestart Malcolm\surnameend},
  \bibinfo{author}{J.~H. \surnamestart Roseboom\surnameend},
  \bibinfo{author}{C.~E. \surnamestart Clark\surnameend} \&
  \bibinfo{author}{W.~\surnamestart Fazar\surnameend} (\bibinfo{year}{1959}):
  \emph{\bibinfo{title}{Application of a technique for research and development
  program evaluation}}.
\newblock {\slshape \bibinfo{journal}{Operations Research}}
  \bibinfo{volume}{7}(\bibinfo{number}{5}), pp. \bibinfo{pages}{646--669},
  \doi{10.1287/opre.7.5.646}.

\bibitemdeclare{article}{Mazzuto15082012}
\bibitem{Mazzuto15082012}
\bibinfo{author}{Giovanni \surnamestart Mazzuto\surnameend},
  \bibinfo{author}{Maurizio \surnamestart Bevilacqua\surnameend} \&
  \bibinfo{author}{Filippo~Emanuele \surnamestart Ciarapica\surnameend}
  (\bibinfo{year}{2012}): \emph{\bibinfo{title}{Supply chain modelling and
  managing, using timed coloured Petri nets: a case study}}.
\newblock {\slshape \bibinfo{journal}{International Journal of Production
  Research}} \bibinfo{volume}{50}(\bibinfo{number}{16}), pp.
  \bibinfo{pages}{4718--4733}, \doi{10.1080/00207543.2011.639397}.
\newblock \eprint{https://doi.org/10.1080/00207543.2011.639397}.

\bibitemdeclare{article}{merlin1974study}
\bibitem{merlin1974study}
\bibinfo{author}{Philip \surnamestart Merlin\surnameend}
  (\bibinfo{year}{1974}): \emph{\bibinfo{title}{A study of the recoverability
  of computer systems}}.
\newblock {\slshape \bibinfo{journal}{Ph. D. Thesis, Computer Science Dept.,
  University of California}}.

\bibitemdeclare{article}{refIPTPN}
\bibitem{refIPTPN}
\bibinfo{author}{F.~\surnamestart Peres\surnameend},
  \bibinfo{author}{B.~\surnamestart Berthomieu\surnameend} \&
  \bibinfo{author}{F.~\surnamestart Vernadat\surnameend}
  (\bibinfo{year}{2011}): \emph{\bibinfo{title}{On the Composition of Time
  {Petri} Nets}}.
\newblock {\slshape \bibinfo{journal}{Discrete Event Dynamic Systems}}
  \bibinfo{volume}{21}(\bibinfo{number}{3}), \doi{10.1007/s10626-011-0102-2}.

\bibitemdeclare{article}{Ramalingam95solvingdifference}
\bibitem{Ramalingam95solvingdifference}
\bibinfo{author}{G.~\surnamestart Ramalingam\surnameend},
  \bibinfo{author}{J.~\surnamestart Song\surnameend},
  \bibinfo{author}{L.~\surnamestart Joscovicz\surnameend} \&
  \bibinfo{author}{R.~E. \surnamestart Miller\surnameend}
  (\bibinfo{year}{1995}): \emph{\bibinfo{title}{Solving Difference Constraints
  Incrementally}}.
\newblock {\slshape \bibinfo{journal}{Algorithmica}} \bibinfo{volume}{23},
  \doi{10.1007/PL00009261}.

\bibitemdeclare{inproceedings}{Zhang2009}
\bibitem{Zhang2009}
\bibinfo{author}{Xiaoling \surnamestart Zhang\surnameend},
  \bibinfo{author}{Qiang \surnamestart Lu\surnameend} \&
  \bibinfo{author}{Teresa \surnamestart Wu\surnameend} (\bibinfo{year}{2009}):
  \emph{\bibinfo{title}{Petri-net based application for supply chain
  management: An overview}}.
\newblock In: {\slshape \bibinfo{booktitle}{IEEM 2009 - IEEE International
  Conference on Industrial Engineering and Engineering Management}},
  \doi{10.1109/IEEM.2009.5373050}.

\end{thebibliography}
\end{document}